\documentclass[12pt]{iopart}

\usepackage{iopams}  
\usepackage{graphicx}
\begin{document}

\title{Slave nodes and the controllability of metabolic networks}

\author{Dong-Hee Kim\footnote[1]{Current address: Department of 
Applied Physics, Helsinki University of Technology, P.O. Box 5100, 
02015 HUT, Finland} and Adilson E. Motter}

\address{Department of Physics and Astronomy and Northwestern Institute on Complex Systems (NICO), Northwestern University, Evanston, IL 60208, USA}
\ead{\mailto{dong.kim@tkk.fi} and \mailto{motter@northwestern.edu}}

\begin{abstract}
Recent work on synthetic rescues has shown that the targeted deletion of specific metabolic genes can often be used to rescue otherwise non-viable mutants. This raises a fundamental biophysical question: to what extent can the whole-cell behavior of a large metabolic network be controlled by constraining the flux of one or few reactions in the network? This touches upon the issue of the number of degrees of freedom comprised by one such network. Using the metabolic network of {\it Escherichia coli} as a model system, here we address this question theoretically by exploring not only reaction deletions but also a continuous of all possible reaction expression levels. We show that the behavior of the metabolic network can be largely manipulated by the pinned expression of a single reaction. In particular, a relevant fraction of the metabolic reactions exhibit canalizing interactions, in that the specification of one reaction flux determines cellular growth as well as the fluxes of most other reactions in optimal steady states. The activity of individual reaction scan thus be used as surrogates to monitor and possibly control cellular growth and other whole-cell behaviors. In addition to its implications for the study of control processes, our methodology provides a new window to study how the integrated dynamics of the entire metabolic network emerges from the coordinated behavior of its component parts.

\end{abstract}

\vspace{2pc}
\noindent{\small Journal-ref: New Journal of Physics 11, 113047 (2009).}

\maketitle

\section{Introduction}

Complex systems are composed of a large number of interacting parts. Physically, this means that the holistic description of a complex system necessarily involves a high-dimensional phase space. Significant previous work on the {\it structure} of complex networks \cite{rev0,rev2,rev1,rev3,rev4} has been based on this paradigm. Herewith we explore an alternative approach, which we argue is appropriate to address the {\it functional} behavior of complex biological systems. By focusing on cellular metabolic networks as a model system of broad significance \cite{lenski,ingram,prl2,stephanopoulos2007,pj,prl1,motter2008}, we show that despite being high dimensional, a metabolic network has a surprisingly small {\it effective} number of degrees of freedom, operationally defined as the number of independent reaction fluxes under steady-state conditions. This indicates that the steady-state dynamics of large complex networks can be significantly more constrained than their structure may suggest. Because the interactions constraining the dynamics also mediate information flow across the network, this result raises the possibility of natural as well as engineered control mechanisms based on the monitoring or
manipulating one or few metabolic reactions.

This study is motivated by the recent discovery that in single-cell organisms the growth defect caused by the deletion of an enzyme-coding gene can often be compensated by the concurrent deletion of a second enzyme-coding gene \cite{motter2008}. Such {\it synthetic rescue} interactions, in which damage compensates for damage, were predicted to even turn some non-viable gene-deficient strains into viable strains after specific additional gene deletions were introduced. Related research has found that cells evolved to optimize growth rate or any other typical function of metabolic fluxes tend to significantly reduce the number of  {\it active} metabolic reactions when compared to typical non-optimal cells \cite{nishi2008}. This spontaneous reaction inactivation explains the role of latent pathway activation and why, sometimes, ``less is more" in cellular metabolism. In particular, it shows that the compensatory perturbations underlying synthetic rescues are generally generated by the inactivation of metabolic reactions that are predicted to be inactive in growth-maximizing states. An outstanding question that stems from these findings is the extent to which the predetermined ``optimal state" activity of a small fraction of reactions can constrain the entire metabolic network to operate close to the corresponding optimal state. Here we investigate this question by considering the flux specification of a single reaction under steady-state conditions.

We focus on the latest reconstructed metabolic network of the bacterium {\it Escherichia coli} K12 MG1655 \cite{feist2007}, which is arguably the most complete {\it in silico} cellular network to date \cite{feist2008}, and we use flux balance-based methods \cite{edwards2000, kauffman2003} to computationally predict functional states of the system (see {\it Appendix A}). Within this framework, we observe that the pinned flux specification of a single reaction, such as the aminodeoxychorismate lyase reaction, can be sufficient to confine the steady-state cellular growth to zero or the maximum possible, without entailing other assumptions. In particular, we predict that the deletion effect of various otherwise essential enzymes, such as enolase (ENO), which catalyzes the conversion of 2-phosphoglycerate into phosphoenolpyruvate, can be compensated (up to a theoretical limit) by the controlled over-expression of a different reaction in the network\footnote{This often requires the coordinated activity of other reactions, as explained in the text.}. To explain and expand on these observations, we first determine how the steady-state condition alone affords such a canalizing interaction between the flux of one reaction and the integrated biomass flux by modeling \textit{master-slave relations} between metabolic reactions.

\begin{figure}
\center{\includegraphics[width=0.8\textwidth]{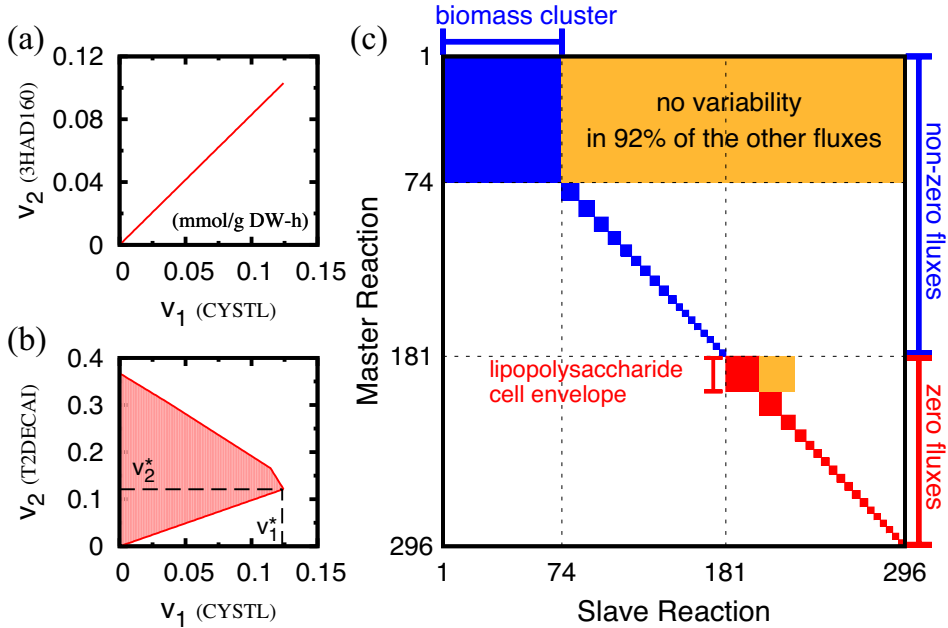}}
\caption{
\label{fig1}
Master-slave relations.  (a) Example of type A relation: fixing flux $v_1$ (master) at any feasible value uniquely determines flux $v_2$ (slave), and vice versa. (b) Example of type B relation: fixing flux $v_1$ (master) at the extreme value $v_1^*$ uniquely determines flux $v_2=v_2^*$ (slave), but the converse does not hold true. The reactions are indicated in parenthesis using the iAF1260 database abbreviation \cite{feist2007}. (c) Clustering characteristics of related fluxes. The diagonal blocks indicate reactions in type A relations, having either nonzero (blue) or zero (red) fluxes in any growth-maximizing state. The off-diagonal blocks (orange) indicate reactions in type B relations, whose fluxes are determined by fixing any of the corresponding block-diagonal reactions at the extreme values of a growth-maximizing state given by the simplex algorithm. The topmost blue block is the biomass cluster, which consists of the biomass reaction itself and 73 other type A-related reactions that uniquely determine the growth rate. Diagonal blocks with less than 4 reactions, which include all fluxes that can be both zero and nonzero at growth-maximizing states, are not shown. Additional information is provided in Table S1.
}
\end{figure}

\section{Flux relations}

We identify two fundamental types of master-slave relations (see figure~\ref{fig1}). We refer to them as {\it type A} relation, in which fluxes $v_i$ and $v_j$ can be uniquely determined by fixing either one of them at any feasible value; and {\it type B} relation, in which flux $v_j$ is uniquely determined in an optimal state where flux $v_i$ is maximized or minimized. These relations define clusters of fully and conditionally coupled fluxes under steady-state conditions, as shown in  figure~\ref{fig1}(c). This figure shows that a total of 73 reactions are in type A relation with the biomass flux, forming a cluster that we refer to as the {\it biomass cluster}.  Notably, 92\% of the other reactions are in type B relation with this cluster for {\it any} growth-maximizing state, i.e., they are ``slaved" by the cluster. This is significant given that the set of growth-maximizing states itself is high dimensional \cite{mahadevan2003,reeds2004}. We also identify a number of smaller clusters. The second largest cluster consists of 22 reactions involved in lipopolysaccharide biosynthesis/recycling and cell envelope biosynthesis, which are nevertheless inactive in all growth-maximizing states. These results are ultimately related to the previously discussed concepts of flux couplings \cite{metatool,burgard2004} and flux correlations \cite{reeds2004,papin2004,vo2004,jamshidi2006}, and thus potentially relevant for the identification of alternatives to known drug targets \cite{duarte2007} and for the prediction of whole-cell metabolic behaviors based on the activity of a small set of reactions \cite{airoldi09}. But what are the underlying mechanisms and functional consequences of this striking canalizing structure?

Physically, the canalizing interactions can be interpreted as a consequence of the steady-state condition: the specification of a certain reaction flux sets constraints on the possible fluxes of other reactions in order to prevent accumulation and depletion of its products and reactants. At the most fundamental level, this is determined by the physical capability of the network to provide alternative pathways to produce and/or consume these compounds. In terms of the region of feasible flux solutions, which is a convex region in the space of fluxes determined by $v_i^{\min}$ and $v_i^{\max}$ ({\it Appendix A}), type A relations are satisfied in the interior of this region and type B relations are satisfied at the borders (cf figures 1(a) and (b)).

\begin{figure}
\center{\includegraphics[width=0.8\textwidth]{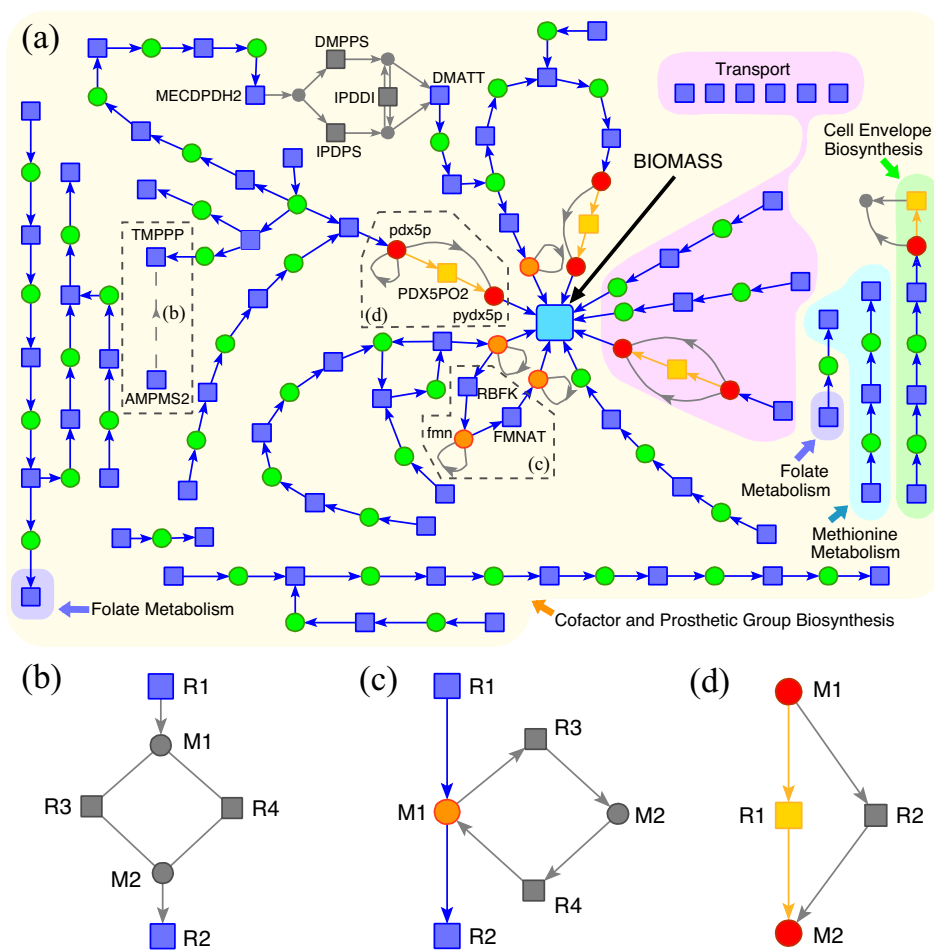}}
\caption{
\label{fig2}
Biomass cluster. (a) Blue boxes and color circles represent the reactions in the biomass cluster and the associated metabolic compounds, respectively. Green circles indicate the metabolites exclusively produced and consumed by reactions in the cluster, while orange and red circles indicate that the metabolite is also produced and/or consumed by other reactions. Reactions and metabolites shown in gray do not belong to the cluster. (b-d) Examples of: (b) reactions in the cluster, such as AMPMS2 and TMPPP, that are coupled through multiple parallel pathways (gray symbols); (c) a metabolite (orange circle), such as fmn, that is shared by a decoupled reaction loop having complete mass balance of the metabolite within the loop; and (d) a reaction (yellow box), such as PDX5PO2, that connects metabolites (red circles) directly linked to the biomass cluster and that is fully coupled to the cluster only in growth-maximizing states. The biomass cluster can be augmented to include $4$ additional such reactions (yellow boxes in (a)). Additional information is provided in Table S2.
}
\end{figure}

Figure \ref{fig2} shows a network representation of the biomass cluster, which is distributed across five functional subsystems of metabolism. Two intriguing properties emerge from this figure. First, despite being fully correlated, the reactions (blue squares) do not form a single connected network, but are instead separated into $16$ different components. This property, which was also suggested in previous work \cite{burgard2004,duarte2007}, can be readily rationalized when different components are linked together by parallel pathways whose combined fluxes are fixed. As shown in the simplified representation of figure~\ref{fig2}(b) for two reactions in the cofactor and prosthetic group biosynthesis, the individual reactions in these  pathways need not to be locked to the activity of the biomass cluster. However,  their combined activity is locked in steady states through the assumed constant concentration of metabolites $M_1$ and $M_2$.

Second, while many reactions in the biomass cluster are coupled through the mass balance of metabolic compounds not involved in other reactions (green circles), we find a number of reactions coupled through metabolites that are also produced and/or consumed by reactions outside the cluster (red and orange circles). This counter-intuitive effect occurs when the reactions in the cluster and the other reactions sharing a common metabolite satisfy mass-balance conditions independent of each other. One such case is shown in figure~\ref{fig2}(c) for the local network of a shared metabolite $M_1$, such as flavin mononucleotide (fmn), where the outside reactions are constrained to produce and consume $M_1$ at exactly the same rate in order to balance metabolite $M_2$. This mechanism can be conceptually understood as the decoupling between two different elementary flux modes \cite{schuster2000,stelling2002}. Figure~\ref{fig2}(d) illustrates a different structure in which a reaction $R_1$, such as pyridoxine 5'-phosphate oxidase (PDX5PO2), is (bidirectionally) coupled to the biomass only when biomass production is maximized. In this case the flux of the parallel reaction $R_2$, which is irreversible, goes to zero in growth-maximizing states, although it is generally nonzero in other states. 

This should be compared with parallel irreversible reactions connecting end-points of the biomass cluster, such as DMPPS and IPDPS in figure~\ref{fig2}(a). While the fluxes of the parallel reactions are not fixed by the biomass production rate in any steady state, their combined flux is, and thus the maximization of the individual fluxes of either DMPPS or IPDPS constrains all the fluxes of the biomass cluster. In the terminology developed above, the cluster is in type B relation with these reactions, i.e., the coupling direction is the opposite of the one shown in figure~\ref{fig1}(c).

\section{Pinned reaction expression}

\begin{figure}
\center{\includegraphics[width=0.9\textwidth]{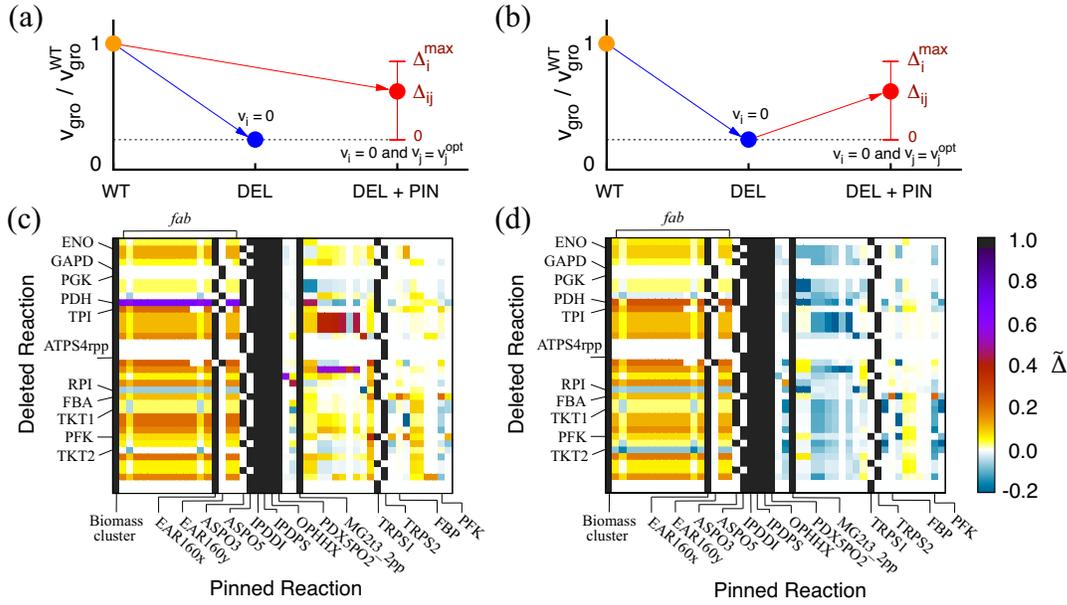}}
\caption{
\label{fig3}
Pinned over- or under-expression of metabolic reactions in reaction-deficient mutants. Two implementations are illustrated, where the WT is assumed to be in a growth-maximizing state and the pinned reaction expression (PIN) is introduced (a) before or (b) after the network response to the reaction deletion (DEL). $\Delta_{ij}$ indicates the growth rate change due to the pinned expression 
of a reaction $j$ following the deletion of a reaction $i$, and $\Delta_{i}^{\max}$ is the theoretical maximum of $\Delta_{ij}$. (c-d) Color-coded normalized growth rate changes $\tilde{\Delta} \equiv \Delta_{ij}/\Delta_i^{\max}$ for the scenarios (a) and (b), respectively. On the horizontal and vertical axes we specify reactions whose pinning significantly increase growth or whose deletions are classified as lethal in experiments \cite{gerdes2003,baba2006,pecdb,book}, respectively. In (c), the data plotted corresponds to mutants having $\Delta^{\max}\ge 10\%$ of the optimal WT growth rate and to pinned expressions having $\tilde{\Delta}>0.3$ for over-expressions ($>0.1$ for under-expressions) for at least one mutant. In (d), the set of mutants and pinned reactions is the same as in (c) to allow direct comparison between the two implementation scenarios. A representation of (d) for data selected according to the same criteria used in (c) is shown in Fig. S1. Additional information is provided in Table S3.
} 
\end{figure}

We now turn to the implications of these mechanisms for the enhancement of biomass production (or growth) in reaction-deficient mutants of {\it E. coli}, which are identical to the wild-type (WT) cells except that one reaction flux is constrained to zero. While the deletion of any reaction in the biomass cluster is lethal, since it forces the biomass flux to zero\footnote{The biomass cluster is a subset of the  previously identified essential reaction core \cite{burgard2004} or lethality core [Uzgil B, unpublished], and it is related to the existence of a medium-independent activity reaction set~\cite{almaas2005}.},  controlling the flux of one such reaction can constrain the system to optimal states. This is achieved {\it without} requiring
the specification of the cell's response to perturbations other than the assumption that the reaction flux can be controlled and the post-perturbed flux distribution reaches a steady state. To extend this analysis to other reactions, we model the changes in cellular growth that may follow a reaction deletion and explore the {\it pinned expression} of a different reaction as a means to restore the growth or to identify surrogates for growth optimization in the reaction-deficient strain (see figure~\ref{fig3}).

Specifically, we test the hypothesis that growth can be restored if the flux of the pinned reaction can be constrained to the optimal flux value identified using flux balance analysis (FBA) in a growth-maximizing state of the reaction-deficient mutant ({\it Appendix A}). In our simulations we assume that the pre-mutation organisms are in growth-maximizing states, as observed in adaptive evolution experiments
\cite{adapt}. We also assume that the early post-mutation state is governed by the minimization of metabolic adjustment (MOMA) hypothesis \cite{segre2002}, which has been shown to describe the flux pattern of deletion mutants and assumes that the perturbed system goes to the closest available steady 
state in terms of Euclidean distances in the space of fluxes. MOMA-predicted fluxes generally correspond to suboptimal states, whose biomass production is lower than the theoretical maximum determined by FBA.

We examine two possible scenarios, corresponding to two different experimental implementations. In the first scenario, we consider that the pinned reaction expression is introduced before the metabolic network responds to the reaction deletion (figure~\ref{fig3}(a)). In the second scenario, we assume that the pinned reaction expression is implemented after the network has responded to the reaction deletion (figure~\ref{fig3}(b)). The first case is modeled as a MOMA-predicted response to the combined perturbation of the reaction deletion and pinned expression (red arrow; figure~\ref{fig3}(a)), while the second case is best modeled as a sequence of two MOMA-predicted responses (blue and red arrows; figure~\ref{fig3}(b)).

Figures \ref{fig3}(c) and (d) show the corresponding growth enhancement predicted for a selection of 38 reaction-deficient mutants. The biomass cluster is preserved for all single-reaction deletions in our simulations, indicating that under steady-state conditions the controlled expression of any of its reactions would tune the metabolic network of the deletion mutant to a growth-maximizing state. 
The same holds true for 11 other reactions identified as directionally coupled to the biomass cluster for a growth-maximizing state identified by the simplex algorithm (i.e., the cluster is type B-related to them), 
namely EAR160x, EAR160y, ASPO3, ASPO5, IPDDI, IPDPS, OPHHX, PDX5PO2, MG2t3\_2pp,
TRPS1, and TRPS2 (see Table S3). Other reactions, such as DMPPS, can be directionally coupled to the biomass cluster for different growth maximizing state. Under the constraints imposed by constant biomass composition and steady-state reaction fluxes, the pinned expression of one such reaction confines the system to an optimal growth rate that depends neither on the MOMA modeling nor on the order of the perturbations (cf figures~\ref{fig3}(c) and (d)).

In addition, we identify reactions not related to the biomass cluster but whose flux pinning would significantly compensate the growth defect of the mutants. For example, while the suboptimal growth rate of the PDH-deficient mutant is predicted to be $0.28$ of the wild-type growth rate, pinning the flux of 3HAD121 increases the normalized growth rate to 0.65, which is $78\%$ of the theoretical maximum (figure~\ref{fig3}(c)). The most significant growth recoveries are found to involve reaction over-expressions ($43$ reaction pairs with $\tilde{\Delta}>0.3$), but the positive impact of single-reaction under-expression \cite{motter2008} was also identified and corresponds to $10$ reaction pairs with $\tilde{\Delta}>0.1$ in figure~\ref{fig3}(c). Although the partial recoveries can vary with the implementation scenario, a statistically significant overlap is observed in figures~\ref{fig3}(c) and (d):
all full recoveries are the same and the growth changes are concurrently positive or  concurrently negative for nearly 78\% of the other cases (see see Figs. S2).

Figure \ref{fig3} includes rescue counterparts for reaction deletions associated with genes identified as {\it essential} in deletion experiments~\cite{gerdes2003,baba2006,pecdb,book}. For example, the over-expression of reactions of the cell envelope biosynthesis associated with {\it fab} genes is predicted to restore growth of several non-viable mutants\footnote{Although our simulations do not predict zero 
suboptimal growth for all these mutants, the predicted growth rates tend to be significantly smaller than the wild-type growth rate.}, 

We have focused on {\it reaction activity} in order to obtain general results entirely determined by the mass balance equations.  However, as the above examples indicate, our approach  also provides useful 
information about {\it gene activity} through known gene-enzyme-reaction relationships~\cite{reed2006}.
This is corroborated by the fact that 36\% of the  enzyme-coding genes in the iAF1260 reconstruction \cite{feist2007} are in one-to-one relationship with metabolic reactions. Moreover, in many cases similar growth rescues can be obtained by pinning the expression of any one out of a large number of different reactions (figure~\ref{fig3}). From the perspective of metabolic control and the identification of alternative
metabolic optimization targets, the latter indicates that one can choose to focus only on a subset of reactions with desirable properties (e.g., one-to-one reaction-gene relation or availability of promoters and markers).

\section{Discussion}

The observation that the metabolic network may be controlled by a small number of degrees of freedom seems to be in accordance with the presence of global regulators \cite{riehl08} as well as the recent experimental observation that few (sometimes far-reaching) mutations can significantly increase the growth rate of {\it E. coli} strains subjected to adaptive evolution \cite{herring2006}. In interpreting our results in the context of metabolic engineering, one should of course not underestimate the experimental 
difficulties involved in the steady-state control of a reaction flux \cite{stephanopoulos2007,stephanopoulos1997}. In particular, the reaction expression is not always correlated with gene expression and the availability of enzymes~\cite{shlomi2007,vazquez2008}.
This is so partly because the balanced activity of a metabolic reaction may depend on the coordinated expression of multiple genes and may be influenced by post-transcriptional effects. While these issues fall outside the scope of this study, we observe however that significant progress has been made in developing expression systems that can lead to tunable reaction expression patterns~\cite{jensen1998,mijakovic2005}. These techniques can expand the applicability of our results as
well as of the recently introduced OptReg platform~\cite{pharkya2006}, which is a versatile framework that exploits reaction down- and up-expression in the overproduction of targeted compounds. 

Specific experimental studies on {\it E. coli} metabolism appear to support our results. Examples of synthetic rescues induced by the suppression of specific metabolic reactions have been discussed in Ref.~\cite{motter2008}, and similar experimental results are also found for reaction over-expressions. For instance, the growth of {\it pgi}-deficient {\it E. coli} mutants fed glucose has been shown to be significantly improved by the over-expression of the soluble transhydrogenase UdhA~\cite{exp5}. This occurs presumably because UdhA restores redox balance when the Pentose Phosphate pathway becomes the primary route of glucose catabolism following the inactivation of phosphoglucose isomerase, which agrees with our predictions. Our modeling of the over-expression of the soluble transhydrogenase reaction leads to 5\% flux increase through the Pentose Phosphate pathway while suppressing the flux through the Entner-Doudoroff pathway.

Other experimental case studies that can be related to this work suggest that the possibility of rescuing a mutant using over-expressions is a general mechanism not limited to {\it E. coli} metabolism. For example, it has been shown that the over-expression of protein PGC-1$\alpha$, a regulator of energy metabolism, promotes the recovery of mitochondrial dysfunction caused by oxidant exposure~\cite{exp1}, apparently by up-regulating mitochondrial biogenesis in tissues with high metabolic demand. A different study has shown that HIV-1 mutants with budding defect are rescued by the over-expression of protein Nedd4-2s~\cite{exp2}, a Nedd4-like ubiquitin ligase of the family recruited by less-complex retroviruses. In humans, the over-expression of mitochondrial valyl tRNA synthetase has been shown to partially rescue cells carrying pathogenic mutations associated with inborn metabolic diseases~\cite{exp3}. Moreover, experiments with insulin receptor-deficient mice, a mutant that develops severe diabetes, indicate that the over-expression of hepatic glucokinase improves glucose tolerance and partially compensates for the metabolic disorders associated with this deficiency~\cite{exp4}. In addition, a number of over-expression rescue interactions have been identified for genes involved in various other cellular functions, most noticeably for yeast~\cite{exp0}, as shown in the Saccharomyces Genome Database~\cite{sgd}. While the mechanisms underlying these examples remain largely unexplained, they highlight the potential significance of the interactions systematically predicted here, particularly for the recovery of lost cellular function and the substitution of known drug and microbial optimization targets.

The reduced effective number of degrees of freedom identified in this study may represent a general property of complex biological networks whose function, like in metabolism, is based on the transport and/or transformation of locally preserved quantities. This includes, for example, food webs and many other resource transportation or transformation networks. In such systems, the suppression of one flux is often accompanied by the enhancement of different fluxes, allowing us to interpret flux over-expressions as reciprocals to the flux down-expressions recently exploited to bypass defective pathways~\cite{motter2008}. We expect that the insights provided by this study will be further expanded in combination with other techniques, such as the extreme pathway~\cite{wilback2002} and elementary flux mode~\cite{schuster2000,stelling2002} analyses, through the study of non-stationary behavior \cite{ing}, and by means of additional applications of network analysis~\cite{rev0,rev2,rev1,rev3,rev4,gallos2007,maslov2007,albert2007,bianconi2008,toro2008}.
Altogether, this promises to improve our understanding of the interaction between the dynamics and the control mechanisms underlying the functional behavior of complex biological networks. 

\ack
The authors thank Natali Gulbahce and Linda Broadbelt for insightful discussions. This work was supported by NSF under Grant No. DMS-0709212 and involved the use of computer equipment funded through the NSF-MRSEC Program DMR-0520513.

\appendix

\section{Network Model and Methods}

The genome-wide {\it in silico} {\it E. coli} metabolic network iAF1260 \cite{feist2007} used in this study consists of $n= 2\, 074$ unique biochemical reactions and $m= 1\, 039$ chemical compounds. The network and the state of the system are conveniently represented by a $m\times n$ matrix of stoichiometric coefficients ${\boldsymbol S}=(S_{ij})$ that indicates the molar proportions of the reactants $i$ in reaction $j$ and a $n$-dimensional vector of reaction fluxes ${\boldsymbol  v}=(v_{i})$, respectively.
Our steady-state analysis is based on identifying solutions for the reaction fluxes through the mass balance equation ${\boldsymbol  S} \, {\boldsymbol  v}={\boldsymbol  0}$ subjected to  $v_i^{\min} \le v_i \le v_i^{\max}$, where $v_i^{\min}$ and $v_i^{\max}$ represent limitations imposed by nutrient availability and thermodynamic, physiological or biochemical constraints \cite{edwards2000}. For concreteness, we consider a limited glucose nutrient environment with maximum uptake rates of $10$ for glucose and $20$ for oxygen (in units of mmol/g DW-h). In glucose aerobic media, $152$ regulated reactions of the reconstructed model are assumed to be inactive and a split of $1$:$1$ is used for the flux ratio between the two  NADH  dehydrogenases~\cite{feist2007}. We focus on the set of $N=1\, 287$ biochemical reactions involving $M=690$ chemical compounds that can be active under these conditions.
Of this total, $853$ reactions are necessarily inactive in growth-maximizing states. Specific metabolic states are determined through two widely used optimization schemes, flux balance analysis (FBA) for growth-maximizing states \cite{edwards2000} and the minimization of metabolic adjustment (MOMA) for post-perturbation states~\cite{segre2002}. FBA finds a solution ${\boldsymbol  v}$ that maximizes the production of biomass, which is represented through an additional reaction in matrix ${\boldsymbol S}$,
and exploits the fact that the biomass flux rate can be mapped to growth rate. MOMA finds a new feasible solution ${\boldsymbol  v}_{new}$ that is closest to the original state ${\boldsymbol  v}_0$ in terms of the Euclidean distance $||{\boldsymbol  v}_{new} - {\boldsymbol  v}_0||$ in the space of fluxes.
The master-slave relations are identified using a variant of flux variability analysis \cite{mahadevan2003}, which identifies the upper and lower bounds of the flux values when the flux of a given reaction is fixed, as in figure~\ref{fig1}(a) and (b). This approach avoids the combinatorial explosion problem~\cite{klamt2002} inherent to extreme pathway analysis~\cite{wilback2002} and elementary flux mode analysis~\cite{schuster2000,stelling2002}, allowing it to be applied to the full-scale network.
The simulations are implemented using the ILOG CPLEX optimization software and the simplex algorithm.

\section{Supplementary Data~\cite{supp}}
\textbf{Table S1.} Reactions in the two largest diagonal blocks indicated in Fig. \ref{fig1}(c).\\
\textbf{Table S2.} Additional information about the biomass cluster in Fig. \ref{fig2}(a).\\
\textbf{Table S3.} Deleted and pinned reactions in Figs. \ref{fig3}(c) and \ref{fig3}(d).\\
\textbf{Figure S1.} Different representation of Fig. \ref{fig3}(d).\\
\textbf{Figure S2.} Direct test of the overlaps between Figs. \ref{fig3}(c) and \ref{fig3}(d).\\
\textbf{Test of different objective functions.} ATP production and lactic acid production.

\end{document}